\documentclass[aps,pra,twocolumn,groupedaddress]{revtex4-2}

\usepackage{graphicx}
\graphicspath{{./figures/}}
\usepackage{amsmath}
\usepackage{braket}
\usepackage{epstopdf}

\begin{document}

\title{Nanosecond-timescale development of Faraday rotation in an ultracold gas}

\author{Jonathan R. Gilbert, Mark A. Watkins and Jacob L. Roberts}
\affiliation{Department of Physics, Colorado State University, Fort Collins, CO 80523}

\email[]{Jon.Gilbert@colostate.edu}


\begin{abstract}
	When a gas of ultracold atoms is suddenly illuminated by light that is nearly resonant with an atomic transition, the atoms cannot respond instantaneously.  This non-instantaneous response means the gas is initially more transparent to the applied light than in steady-state.  The timescale associated with the development of light absorption is set by the atomic excited state lifetime.  Similarly, the index of refraction in the gas also requires time to reach a steady-state value, but the development of the associated phase response is expected to be slower than absorption effects.  Faraday rotation is one manifestation of differing indices of refraction for orthogonal circular light polarization components.  We have performed experiments measuring the time-dependent development of polarization rotation in an ultracold gas subjected to a magnetic field.  Our measurements match theoretical predictions based on solving optical Bloch equations.  We are able to identify how parameters such as steady-state optical thickness and applied magnetic field strength influence the development of Faraday rotation.	   
\end{abstract}


\maketitle

\section{Introduction}
\label{sec:Intro}
 
	Extensive theoretical and experimental research has been performed studying near-resonant light interacting with ultracold atomic gases, ranging from dilute (e.g. low spatial number density) \cite{PhysRevLett.91.223904,PhysRevLett.117.073002} to high density ensembles \cite{PhysRevA.87.053817,PhysRevA.93.063835,PhysRevA.96.053629}.  A description using coupled dipoles adequately captures physics in a low density gas \cite{PhysRevLett.117.073003,doi:10.1080/09500340.2017.1380856}.  Accurate predictions for light in high density gases, where atom-atom interactions become relevant, are also being pursued \cite{Jones_2016,PhysRevA.96.033835}.  Identifying subtleties linked to the physics in these systems is ongoing.  For instance, recent theoretical calculations indicate that transitioning from a scalar description to one that includes the vector nature of light (i.e. polarization) can significantly alter predictions of phenomena such as Anderson localization of light \cite{PhysRevLett.112.023905,PhysRevA.90.063822}.  While fundamental studies of light and its interactions with matter date back centuries, it is clear there are still open questions highlighted by discrepancies between current theoretical models and experimental results \cite{PhysRevA.97.053816}.  The important role resonant light plays in a wide variety of fields, including quantum simulation \cite{RevModPhys.86.153}, precision spectroscopy \cite{PhysRevLett.94.153001}, optical clocks \cite{PhysRevLett.123.033201}, and ultracold plasmas \cite{Lyon_2016} to name a few, encourages the continued study of near-resonant light interactions.  In this article, we examine how a phase-associated effect, Faraday rotation, develops in concert with absorption as atoms in a dilute ultracold gas transition from a state of being transparent to being optically thick.   
	    
	The characteristic response time of a gas of atoms to light that is near-resonant or resonant with a particular transition is determined by the atomic excited state lifetime of that transition.  One implication of this is that if near-resonant or resonant light is suddenly applied to a gas of atoms, the gas will be effectively transparent until the atoms have enough time to develop an appreciable dipole response to the light.  This has been demonstrated theoretically and experimentally, for example, in measurements of optical precursors \cite{PhysRevLett.96.143901}, optical free induction decay \cite{PhysRevA.56.1564}, and related effects \cite{PhysRevA.84.011401}.  However, these measurements have focused on absorption effects.  The atom gas can also shift the phase of incident light (i.e. have a real component of an index of refraction), but similar to absorption, cannot do so instantaneously.  Since the phase response is non-instantaneous, related polarization effects such as Faraday rotation will also take time to develop.  
	
	At first glance, the timescale for the phase shifts that underly the phase effects might be expected to be approximately twice as long as for absorption effects since the phase effects manifest themselves linearly with the electric field while absorption effects are observed via light intensity (proportional to the electric field squared).  Through the work presented in this article, we find that the associated physics in a realistic system is more complicated than a straightforward ratio of two relationship.  We have conducted experiments measuring the time development of Faraday rotation in an optically thick ultracold gas and then compared those experimental results with theoretical predictions.  Aside from investigating the associated basic physics, these considerations are relevant if sufficiently short pulses are used in situations where phase shifts are important, as can be the case in cavity QED \cite{Cimmarusti_2013} and interacting Rydberg gases \cite{Firstenberg2013}.
	
	Our experiments consisted of suddenly turning on a resonant linearly polarized laser beam through a gas of ultracold atoms in a magnetic field and monitoring the intensity and polarization of the transmitted light as a function of time.  After a sufficiently long period of time, the system reaches quasi-steady state.  We refer to this as a quasi-steady state because there is a period in time where the light transmission stops changing rapidly with time, but there is then a slow optical pumping that works to bring the gas to a true steady-state over a longer timescale.  Along with a significant absorption of the light determined in part by the atom density, there is a modification of the ellipticity and direction of the light polarization.  The latter effect can be understood as being due to different values of the real part of the index of refraction of the right-handed and left-handed circular polarization components of the incident light.  By measuring the development of the intensity and polarization of the light as a function of time, the time-dependent absorption and phase shifts of the light in the gas can be characterized and compared to theoretical expectations.  Our experimental data is in good agreement with calculations based on optical Bloch equations.

	In the rest of this article, we describe the theoretical modeling of our system and present results based on experimental measurements. Section \ref{sec:Theory} presents the theory used to model the time-dependent evolution of the light in an ultracold gas.  Section \ref{sec:Experiment} describes our experimental implementation used to perform measurements of the time-dependent development of Faraday rotation in an ultracold gas. Section \ref{sec:PhaseResponse} presents the results based on experiments and discusses various influences on the response times observed. Lastly, section \ref{sec:Conclusions} summarizes our conclusions.

\section{Theory}
\label{sec:Theory}
	
	We use Maxwell's equations and a set of optical Bloch equations based on our experimental conditions to calculate the time-dependent transmitted light intensity and polarization.  The spatial extent of the gas as compared to the beam size and wavelength of the incident light is such that diffraction effects are negligible and thus not included in this treatment.  One advantage of this is that we can model the gas as having a uniform spatial density of atoms rather than having to model the density variations that exist in the actual experiment. We also ignore the finite speed of light with regard to the propagation of light intensity changes through the ultracold gas, which is a reasonable approximation for our conditions.   

\begin{figure}[htbp]
\centering
\includegraphics[width=\linewidth]{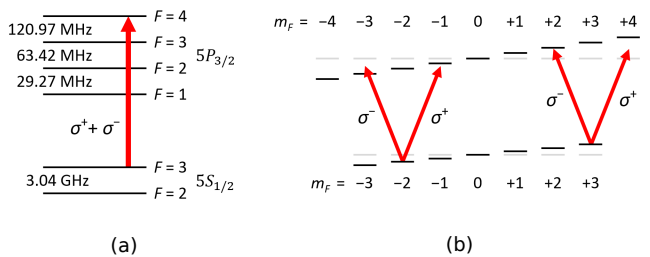}
\caption{\label{fig:D2Line} Relevant energy levels for our theoretical calculations and experimental measurements of the time-dependent Faraday rotation of light in an ultracold gas.  The incident light is composed of equal parts $\sigma^+$ and $\sigma^-$ circular polarization and is represented by the red arrow(s). Part (a) shows the hyperfine structure in the D2 (780 nm) line in $^{85}$Rb (the ground state \cite{Barwood1991} and excited state \cite{Das_2008} splittings are not to scale).  The incident light is primarily resonant with the $5S_{1/2}$ $F=3$ to $5P_{3/2}$ $F=4$ cycling transition. Part (b) depicts the magnetic sublevels in the $5S_{1/2}$ $F=3$ ground state and $5P_{3/2}$ $F=4$ excited state.  The gray lines represent the degenerate (no magnetic field) magnetic sublevels and the black lines represent the Zeeman-shifted sublevels.  Two examples of the $\sigma^+$ and $\sigma^-$ polarization components with respect to particular magnetic sublevel transitions are shown with different amounts of detuning resulting from the energy shifts.}
\end{figure}

	To model our experimental measurements, we use parameters appropriate to the D2 line for $^{85}$Rb (shown in figure \ref{fig:D2Line}(a)) for light that is nearly resonant on the $F=3$ to $F=4$ cycling transition.  The incident light is linearly polarized and propagates along the direction of an applied magnetic field, so it is natural to consider the light as being composed of equal parts $\sigma^+$ and $\sigma^-$ circular polarization components.  The applied magnetic field produces Zeeman shifts across the magnetic sublevels which causes the $\sigma^+$ and $\sigma^-$ polarization components to become frequency detuned by different amounts for transitions that share the same ground state magnetic sublevel (see figure \ref{fig:D2Line}(b)).  Accounting for the range of induced detunings along with the atoms' relative transition strengths in the calculation captures both the absorption and the phase shift associated with the $\sigma^+$ and $\sigma^-$ polarization components.  If the response of the multi-level atoms subjected to the magnetic field leads to a differential phase shift between the polarization components then the result will be a polarization rotation of the light.
	
\subsection{The polarization components}
\label{sec:Efield}
	
	Our theoretical treatment begins with a linearly polarized plane wave propagating in the $\hat{z}$-direction.  The plane wave, $\vec{E}(z,t)$, is incident on a gas of effectively stationary atoms, where $z$ is the spatial coordinate and $t$ is time. There is assumed to be no spatial variation in the directions perpendicular to the direction of propagation.  Starting with Maxwell's equations and the plane wave assumption leads to the wave equation  

\begin{equation}
\label{eq:WaveEq}
\frac{\partial^2 \vec{E}(z,t)}{\partial z^2} = \mu_0 \frac{\partial^2 \vec{D}(z,t)}{\partial t^2},
\end{equation}

\noindent
where $\vec{D}(z,t) = \epsilon_0 \vec{E}(z,t) + \vec{P}(z,t)$, $\mu_0$ is the vacuum permeability, $\epsilon_0$ is the vacuum permittivity and $\vec{P}(z,t)$ is the polarization response of the ultracold gas.  The plane wave solution is expressed as 

\begin{equation}
\label{eq:TotE}
\vec{E}(z,t) = \tilde{E}_{+}(z,t)\hat{\sigma}_{+} + \tilde{E}_{-}(z,t)\hat{\sigma}_{-},
\end{equation}

\begin{equation}
\label{eq:Ecomp2}
\tilde{E}_{\pm}(z,t) = \tilde{A}_{\pm}(z,t)e^{(ik_0z-i\omega t)},
\end{equation}

\noindent
where $\hat{\sigma}_{+}$ and $\hat{\sigma}_{-}$ are circular basis unit vectors and the $\pm$ subscript corresponds to the $\sigma^+$ and $\sigma^-$ circular polarization components, respectively.  $\tilde{A}_{\pm}(z,t)$ are the polarization amplitude and phase components of the wave, $k_0$ is the vacuum wave number, and $\omega$ is the optical frequency.  The relative phase between the $\sigma^+$ and $\sigma^-$ polarization components of the incident light is initially set to be zero.  The atoms' polarization response is expressed as 

\begin{equation}
\label{eq:TotP}
\vec{P}(z,t) = \tilde{P}_{+}(z,t)\hat{\sigma}_{+} + \tilde{P}_{-}(z,t)\hat{\sigma}_{-},
\end{equation}

\begin{equation}
\label{eq:Pcomp}
\tilde{P}_{\pm}(z,t) = \frac{\epsilon_0}{k_0}\tilde{\beta}_{\pm}(z,t)e^{(ik_0z-i\omega t)},
\end{equation} 

\noindent
where $\tilde{\beta}_{\pm}(z,t)$ are complex amplitudes corresponding to the $\sigma^+$ and $\sigma^-$ polarization component dipole responses of the atoms.  Steady-state treatments generally express the atoms' polarization response as $\vec{P} = \epsilon_0 \chi \vec{E}$, where the susceptibility, $\chi$, is a constant.  This is correct once the system has reached steady-state, but it is not an applicable expression for our calculations since we are interested in what happens in the system while the atoms' polarization response is still developing with time.

	To calculate the effect these dynamics have on the light in the gas, we derive an envelope equation by substituting \eqref{eq:TotE} and \eqref{eq:TotP} into \eqref{eq:WaveEq}.  We apply the slowly-varying-envelope-approximation and $\frac{\partial \tilde{P}_{\pm}}{\partial t} << \omega \tilde{P}_{\pm}$, which leads to 

\begin{equation}
\label{eq:EnvEq}
\frac{\partial \tilde{A}_{\pm}(z,t)}{\partial z} = \frac{i}{2}\tilde{\beta}_{\pm}(z,t).
\end{equation} 

	A key feature of \eqref{eq:EnvEq} is that it captures the time-dependent evolution of the atoms' dipole responses across the $\hat{z}$-direction spatial extent of the gas.  To better appreciate the importance of including the spatial extent in the calculation, it is useful to first examine the evolution of near-resonant light interacting with an optically thin gas of atoms in a simplified system as compared to the real $^{85}$Rb states.     
	
\subsection{Atomic dipole response}
\label{sec:Simple}
	
	A simple $F=0$ to $F=1$ transition can be used to examine the relevant general physics of driving an optically thin gas with light composed of equal parts $\sigma^+$ and $\sigma^-$ polarization components.  An applied magnetic field induces Zeeman shifts in the excited state magnetic sublevels that causes the incident light polarization components to become detuned by an equal and opposite amount with magnitude $\lvert\delta\rvert$ from the zero magnetic field transition resonance.  The optical Bloch equation for the dipole coherence of the $\Delta m_F =+1$ ground-excited state transition in a rotating frame can be expressed as
	
\begin{equation}
\label{eq:SimpleOBE}
\dot{\rho}_{1,2} = -(i\delta+\gamma/2)\rho_{1,2}+iA_{+}'(\rho_{1,1}-\rho_{2,2}),
\end{equation}

\noindent	
where the subscripts $1$ and $2$ denote the $m_F=0$ ground state and $m_F=+1$ excited state magnetic sublevels respectively. $\gamma$ is the damping rate (i.e. inverse excited state lifetime of the transition), and $A_{+}'$ corresponds to the $\sigma^+$ polarization component of the incident light.  The light is assumed to have an instantaneous turn-on and is assumed to be very low-intensity ($A_{+}' \ll \gamma$) so that the excited state population remains ignorable.  Given the initial condition, $\rho_{1,2}=0$ at $t=0$, an analytic solution for \eqref{eq:SimpleOBE} can be expressed as

\begin{equation}
\label{eq:SimpleSolution}
\rho_{1,2} = \frac{2A_{+}'}{\gamma}\frac{2\delta/\gamma+i}{1+(2\delta/\gamma)^2}\left[1-\exp\left(-\frac{\gamma}{2}(1+2i\delta/\gamma)t\right)\right].
\end{equation}	
	
\begin{figure}[htbp]
\centering
\includegraphics[width=\linewidth]{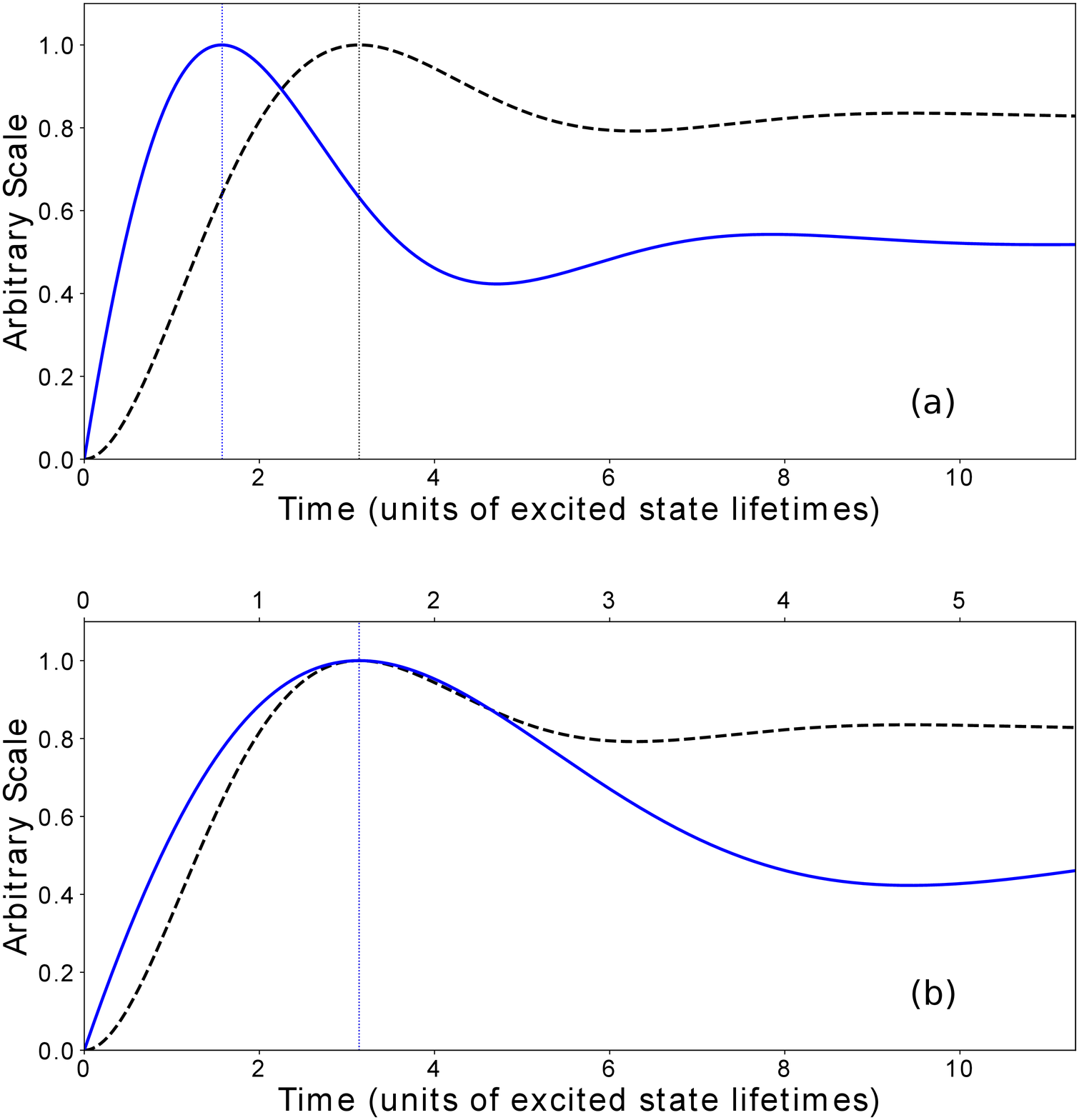}
\caption{\label{fig:SimpleShift} Time-dependent optical depth (blue solid curve) and time-dependent phase (black dashed curve) of the transmitted $\sigma^+$ polarization component through an optically thin gas of atoms driven on a simple $F=0$ to $F=1$ transition, where $A_{+}'=.01\gamma$ and $\eta=1$.  The magnitude of the induced detuning in the calculations is $\lvert\delta\rvert = 1.0 \gamma$.  The blue (black) dotted vertical line indicates the time when the optical depth (phase response) reaches its maximum.  Subfigure (a) shows the curves plotted on the same horizontal axis, with a clear separation between the times corresponding to the maximum of each curve. Subfigure (b) shows that when the phase response is plotted with respect to the bottom horizontal axis and the optical depth is plotted with respect to a separate horizontal axis scaled by a factor of 2 (shown on top of the subfigure), the dotted vertical lines exactly overlap.  Note that the vertical axis is a normalized scale.}
\end{figure}

	The imaginary part, $\operatorname{Im}(\rho_{1,2})$, is associated with absorption while the real part, $\operatorname{Re}(\rho_{1,2})$, is associated with an index of refraction.  Working in normalized units for the electric field amplitude, the transmitted $\sigma^+$ light intensity through the optically thin gas can be characterized by an optical depth (i.e. number of $e^{-1}$ absorption lengths) that is expressed as
	
\begin{equation}
\label{eq:SimpleOD}
\mathrm{O.D.}_+ = -\ln\left([1-\eta\operatorname{Im}(\rho_{1,2})]^2\right),
\end{equation}        

\noindent
where $\eta$ is a unitless scale factor introduced to account for the gas density and physical constants relating the dipole coherence to the light absorption.  For the calculations in this section, we choose $\eta$ such that $\mathrm{O.D.}_+ \sim .01$ for $t \gg 1/\gamma$.  Figure \ref{fig:SimpleShift}(a) shows \eqref{eq:SimpleOD} and $\eta\operatorname{Re}(\rho_{1,2})$ plotted as functions of time with the same horizontal time axis. Figure \ref{fig:SimpleShift}(b) shows the same curves, but with the optical depth plotted with an additional time axis that is a factor of 2 shorter, where the new time axis is displayed on the top of the figure.  In this case, the amplitude and phase component differ by exactly a factor of 2 with regard to their peak response.  A Taylor expansion of $\eqref{eq:SimpleSolution}$ about $t=0$ shows the leading order for the amplitude component is linear in time and the leading order for the phase component is quadratic in time.  This produces the difference in curvature visible at early times in figure \ref{fig:SimpleShift}.
	
\begin{figure}[htbp]
\centering
\includegraphics[width=\linewidth]{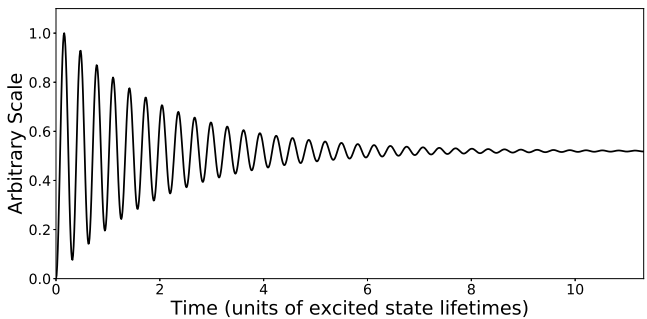}
\caption{\label{fig:PhaseOsc} Phase response for an optically thin gas of atoms driven on a simple $F=0$ to $F=1$ transition with a large detuning, $\lvert\delta\rvert = 20\gamma$.}
\end{figure}

	We note that this simple calculation also provides insight into the evolution of the index of refraction of an absorptive medium illuminated by a very short (with respect to the medium's characteristic radiative lifetime) pulse of light.  For a sufficiently large detuning, the phase response of the medium oscillates rapidly about an average nonzero value, as shown in figure \ref{fig:PhaseOsc}.  Observation timescales that average over this oscillation period will measure an appreciable index of refraction that develops in a very short amount of time, even for sub-lifetime timescales.  In other words, for light detuned much more than a natural linewidth from an atomic transition, the apparent index response timescale will be given by the inverse of the detuning for measurements with sufficiently coarse time sensitivity.         	

\begin{figure}[htbp]
\centering
\includegraphics[width=\linewidth]{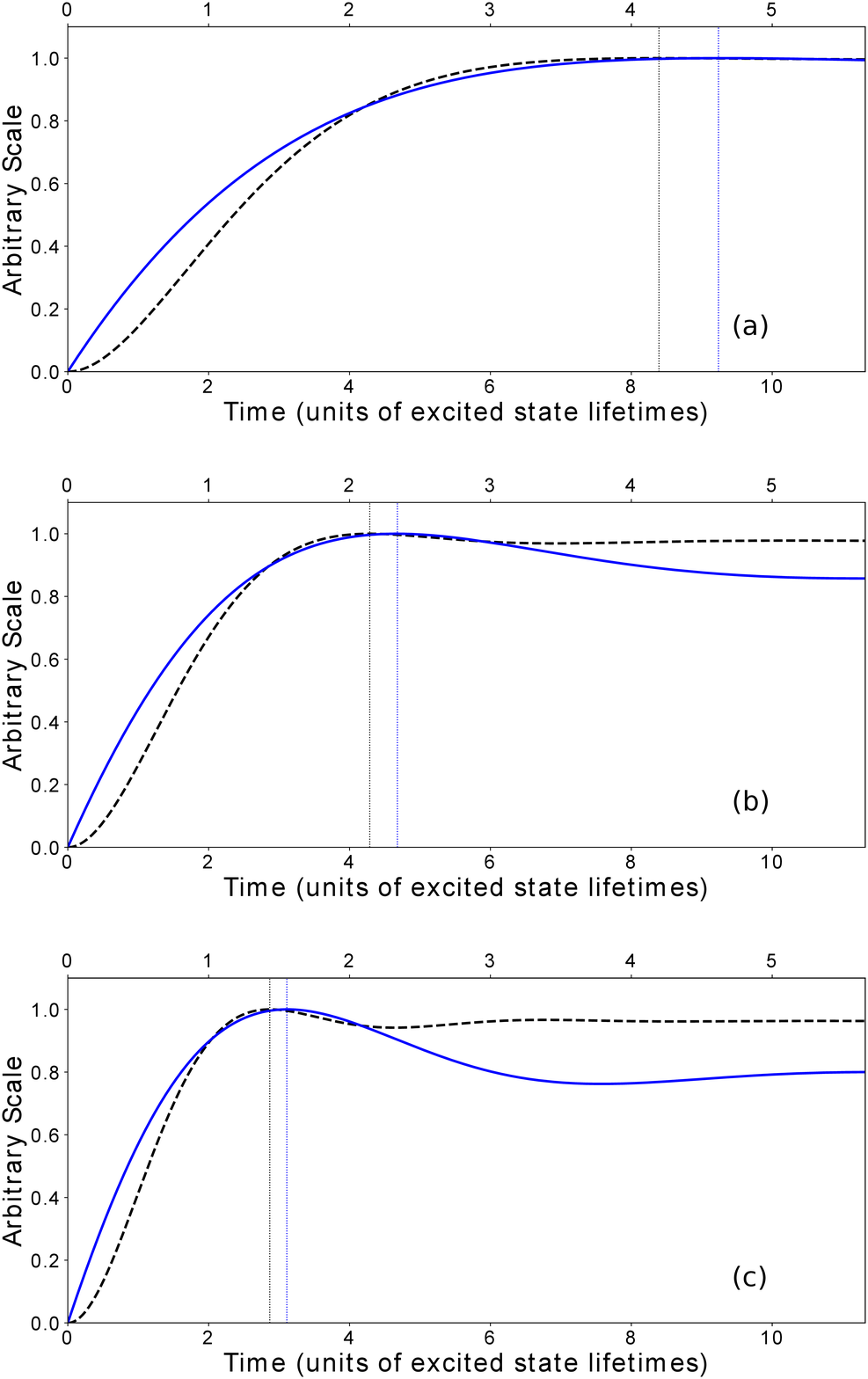}
\caption{\label{fig:SimpleCurves} Time-dependent optical depth (blue solid curve) and time-dependent phase (black dashed curve) of the transmitted $\sigma^+$ light through an optically thin gas of multi-level atoms driven on an $F=3$ to $F=4$ transition in the presence of increasing applied magnetic field strengths.  The ground state magnetic sublevels have initial populations that are evenly distributed.  The optical depth is plotted with respect to the top horizontal axis and the phase response is plotted with respect to the bottom horizontal axis.  The subfigures (a)-(c) are the results when the magnitude of the magnetic field induced detuning for the outermost allowed transition between ground and excited state magnetic sublevels is (a) $\lvert\delta\rvert = 0.5 \gamma$, (b) $\lvert\delta\rvert = 1.0 \gamma$, and (c) $\lvert\delta\rvert = 1.5 \gamma$.  The blue (black) dotted vertical line indicates the time when the optical depth (phase response) reaches its maximum.}
\end{figure}

	Extending the simple calculation to a transition with an $F>0$ ground state results in an immediate departure from a factor of 2 difference between the amplitude and phase components, however.  To illustrate this, we use the $F=3$ to $F=4$ transition with magnetic sublevels depicted in figure \ref{fig:D2Line}(b).  The additional magnetic sublevels means there is a range of Zeeman induced detunings associated with the various transitions.  A transition at the edge of the sublevels (e.g. $m_F=+3$ to $m_F=+4$) has the largest detuning labeled $\delta$.  Assuming low-intensity incident light with an instantaneous turn-on, the optically thin gas can be treated by solving 14 independent optical Bloch equations for the dipole coherences.  The complex amplitude of the $\sigma^+$ ($\sigma^-$) polarization component dipole response is then proportional to the superposition of dipole coherences corresponding to $\Delta m=+1$ ($\Delta m=-1$) transitions.  The dipole coherences have solutions of the form expressed by \eqref{eq:SimpleSolution}, but with a detuning that is different for each transition.  Figure \ref{fig:SimpleCurves} shows that the additional frequency components lead to less than a factor of 2 between the peak response times for the amplitude and phase.  As the magnetic field induced detuning increases there is a faster response for both the amplitude and the phase.   
	
	In contrast to these simple systems, our experiments were not performed in an optically thin gas, but rather an optically thick gas.  This plays a role in the predictions of the total light transmission through the gas and requires including the $\hat{z}$-direction absorption of the ultracold gas in the calculations.  We describe this inclusion in the next section.

\subsection{Dividing the gas into increments}
\label{sec:Increments}
	  
	We return to the envelope equation derived in section \ref{sec:Efield} with the motivation of including the $\hat{z}$-direction absorption to accurately model the time-dependent evolution of light in an optically thick ultracold gas.  The general solution to (\ref{eq:EnvEq}) with respect to an initial reference position, $z=0$, can be expressed as 

\begin{equation}
\label{eq:Acomp}
\tilde{A}_{\pm}(z,t) = \tilde{A}_{\pm}(0,t) + \frac{i}{2}\int_{0}^{z}\mathrm{d}z'\tilde{\beta}_{\pm}(z',t).
\end{equation}

	We have approximated the gas as having a uniform density, which is reasonable in the absence of diffraction effects.  Our theoretical calculations depend on numerically integrating \eqref{eq:Acomp}.  To do so, it is necessary to identify the contributions to the dipole response terms in the integrand.  We obtain the dipole response terms in the integrand by first calculating the density matrix for all states associated with the $5S_{1/2}$ $F=3$ to $5P_{3/2}$ $F=4$ transition in $^{85}$Rb. We include the magnetic sublevels of the ground state ($F=3$) and excited state ($F=4$) shown in figure \ref{fig:D2Line}(b).  Time-dependent Faraday rotation of the total light is introduced into the calculation by including a magnetic field in the $\hat{z}$-direction.  This incorporates the Zeeman induced detunings associated with the energy level shifts and leads to 256 coupled optical Bloch equations that we solve to determine a 16 $\times$ 16 density matrix 
	
\begin{equation} 
\label{eq:DensMatrix}
\rho = 
\begin{pmatrix}
\rho_{1,1} & \dots & \rho_{1,16} \\
\vdots & \ddots & \vdots \\
\rho_{16,1} & \dots & \rho_{16,16}
\end{pmatrix},
\end{equation}
 		 
\noindent 
where the subscripts 1 to 7 correspond to the magnetic sublevels ($m_F = -3,...,+3$) in the ground state and 8 to 16 correspond to the magnetic sublevels ($m_F = -4,...,+4$) in the excited state.

	Once the density matrix is determined, we calculate $\tilde{\beta}_{+}(z,t)$ by summing the dipole coherence terms (i.e. off-diagonal elements) that correspond to $\Delta m_F = +1$ ground-excited state transitions and $\tilde{\beta}_{-}(z,t)$ by summing the dipole coherence terms that correspond to $\Delta m_F = -1$ ground-excited state transitions. Besides coherences between ground and excited states, there are also ground-ground (e.g. $\rho_{1,2}$) and excited-excited coherences (e.g. $\rho_{8,9}$), and they play different and lesser roles in the overall response of the atoms.

	A critical aspect of our calculations is to divide the ultracold gas into equally spaced increments of equal optical depth along the $\hat{z}$-direction.  We calculate the density matrix for each increment at each time step, where the density matrix for an increment is used to calculate the average dipole response of the atoms within that increment.  The local driving field for subsequent increments is then the superposition of the incident driving field and the dipole responses from preceding increments.  The polarization components of the local driving field for an increment located at position $z$ are given by \eqref{eq:Acomp}, where the position of the increment with respect to $z=0$ corresponds to the limit of integration.  
	
	For our calculations, we use 20 increments with each increment having an optical depth of 1/20 the total optical depth, where the total optical depth is typically between 1 and 2.  Once the complex amplitudes of the polarization components are calculated for each increment at each time step, we can determine the time-dependent phase difference of the transmitted light	 
		
\begin{equation}
\label{eq:Phasediff}
\Delta\phi(t) = \phi_{+}(t) - \phi_{-}(t),
\end{equation}

\noindent 
where $\phi_{\pm}(t)$ are the phases associated with the complex polarization components of the light given by (\ref{eq:Acomp}).  Equation \eqref{eq:Phasediff} represents the phase difference between the $\sigma^+$ and $\sigma^-$ polarization components of the total transmitted light after the light has propagated through the full $\hat{z}$-direction spatial extent of the gas.  The amount of polarization rotation incurred by the light as it propagates through the gas is directly attributable to the phase difference. 
	
\subsection{Calculation parameters}
\label{sec:Parameters}

	To make theoretical predictions that correspond to our experimental conditions, we use experimental data to constrain the physical parameters required for the calculations.  These parameters include the magnitude of the magnetic field, the laser detuning, and the average initial ground state magnetic sublevel population distribution (referred to as ``m-state distribution" for the remainder of this article) in the ultracold gas.  The acousto-optic modulator (AOM) that we use to turn the incident light on quickly induces a linear frequency chirp during turn-on.  We independently measured that to be the case and so include a linear chirp when calculating the atom response.  

	Theoretical predictions and experimentally measured transmission data collected over a range of experimental conditions are used to perform a least-squares-minimization to find best-fit values for these parameters.  The calculations maintain no assumptions of low-intensity or steady-state.  This means that all 256 coupled differential equations required to determine the density matrix must be solved for each set of optimization parameters.  Furthermore, this number of equations must be solved for each increment at each timestep, equating to over 5000 coupled equations being solved at each timestep.
	
	Through auxiliary measurements at larger detuning and examining all measurements across multiple conditions, we sensitively determined the light detuning and magnetic field calibrations.  The m-state distribution varied from day-to-day, and so fit parameters were introduced for each day's data.  It was computationally efficient and sufficient to just capture the gross features of the m-state distribution, so the fit parameters were the coefficients of a second-order polynomial with variable $m$ and the m-state distribution was assigned based on that polynomial.  The net result of these calculations are predictions of the time-dependent light in an ultracold gas of atoms corresponding to our experimental conditions.
		
\section{Experiment}
\label{sec:Experiment}

\begin{figure}[htbp]
\centering
\includegraphics[width=\linewidth]{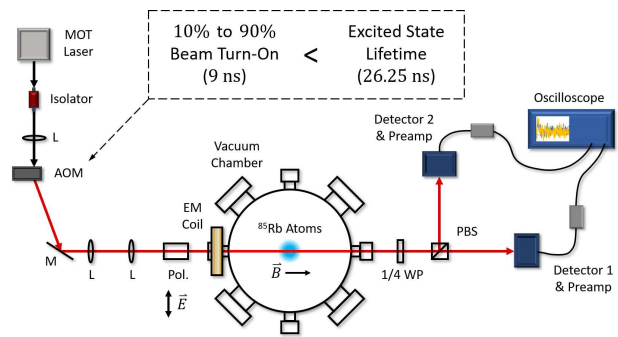}
\caption{\label{fig:Exp} Schematic depicting the experimental set-up used to measure the time-dependent Faraday rotation in an ultracold gas.  The beams used to form the initial MOT are left out for clarity.  Note that the near-resonance beam has a turn-on time that is faster than the excited state lifetime of the atoms.}
\end{figure}

	Our experimental measurements of the time-dependent development of Faraday rotation were performed using an ultracold gas of $^{85}$Rb.  A near-resonance laser beam tuned to the $5S_{1/2}$ $F=3$ to $5P_{3/2}$ $F=4$ cycling transition was turned on rapidly over a timescale faster than the excited state lifetime ($\tau=26.25$ ns \cite{doi:10.1063/1.2035727}) of the atoms. This beam was directed through the center of the ultracold gas.  The total transmitted light was decomposed into orthogonal polarization components and detected on two independent fast photoreceivers.  A schematic depicting the experimental implementation is shown in figure \ref{fig:Exp}.

	We started our experiments by forming an $^{85}$Rb Magneto-Optical Trap (MOT) using standard techniques \cite{PhysRevLett.59.2631}.  After the MOT formation, the trap light and magnetic fields were turned off and the atoms were given 4 ms to expand.  This resulted in an ultracold gas with a root-mean-square (RMS) spatial extent in one dimension of approximately 0.9 mm (determined through separate absorption imaging measurements).  A magnetic field was then applied to the gas.  The magnitude of the magnetic field at the location of the atoms was set to a value between 1.3 Gauss and 6.2 Gauss. We did not increase the magnetic field further in this iteration of the experiment because doing so results in the development of a non-negligible dipole amplitude associated with the $5S_{1/2}$ $F=3$ to $5P_{3/2}$ $F=3$ transition.  Including this transition along with possible decay paths substantially increases the number of coupled differential equations required to accurately model the system. 
	
	With a well-defined axis provided by the magnetic field, we applied a near-resonance laser beam along the same direction through the gas.  The near-resonant beam was derived from the MOT laser and a rapid turn-on time was realized by tightly focusing the beam into a 200 MHz AOM.  The AOM driver was triggered by a 5 ns rise time function generator which resulted in a 10$\%$ to 90$\%$ 1st order deflection turn-on time of 9 ns.  The deflected beam was collimated and sized to a spot size of 364 $\mu$m, which led to a peak intensity polarization-averaged saturation parameter of $I/I_{sat}=$ 0.2, where $I_{sat}=$ 4.8 mW cm$^{-2}$.  The fast photoreceivers used for detection resulted in a signal-to-noise ratio that limited our ability to reduce the intensity much lower without dramatically increasing the amount of data required for a measurement.  The near-resonance beam was passed through a Glan-Thompson polarizer external to the vacuum chamber so the incident light on the atoms was linearly polarized perpendicular to the magnetic field direction.
	
	Positioned on the output side of the vacuum chamber was a quarter waveplate followed by a polarizing beam splitter (PBS) cube.  The orientation of the quarter waveplate's fast axis with respect to the input field's polarization direction determined what type of signal we measured.  The two primary orientations we used for our measurements were 45$^{\circ}$ and 0$^{\circ}$.  Using the 45$^{\circ}$ orientation led to the $\sigma^+$ and $\sigma^-$ polarization components of the total transmitted light being split into separate paths after the PBS.  In this configuration, we measured the transmitted light with and without ultracold atoms present in the vacuum to determine the optical depth associated with each of the polarization components.  
	
	Using the 0$^{\circ}$ orientation and having no ultracold atoms present in the vacuum led to the incident light being transmitted through the PBS onto detector 1 (see figure \ref{fig:Exp}), and only a small background signal on detector 2.  When atoms were present, any relative phase shift between the $\sigma^+$ and $\sigma^-$ polarization components imparted by the atoms while responding to the incident light caused the polarization vector of the total light to rotate, resulting in a time-dependent transmission signal developing on detector 2.  
	 
\begin{figure}[htbp]
\centering
\includegraphics[width=\linewidth]{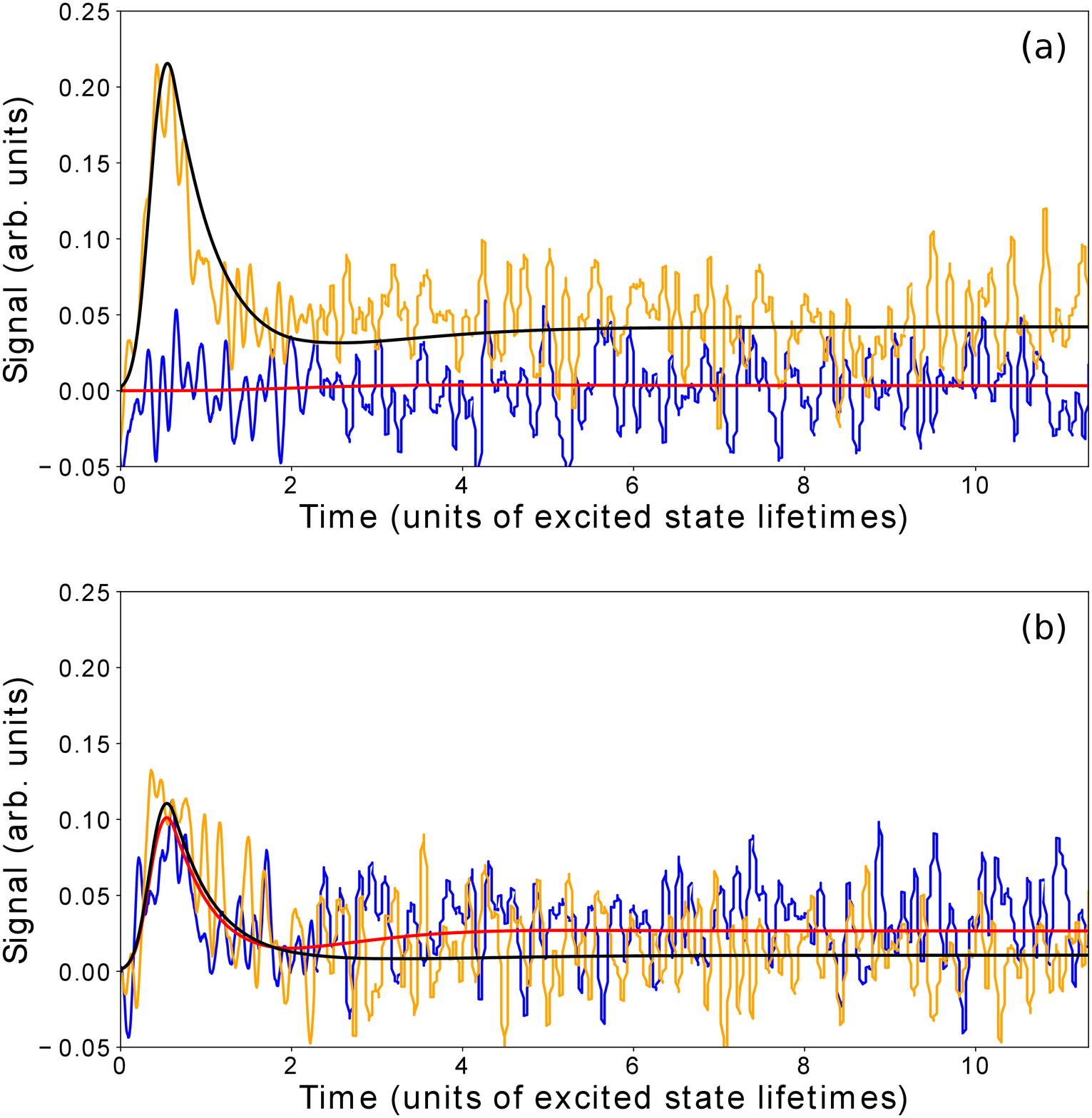}
\caption{\label{fig:Data13} Experimental data and theoretical calculations in the presence of a $B = 1.3$ Gauss magnetic field along the axis of light propagation. Subfigure (a) is data and predictions when the quarter waveplate's fast axis is positioned at 0$^{\circ}$, and subfigure (b) is data and predictions when the quarter waveplate's fast axis is positioned at 45$^{\circ}$.  The yellow and blue data are measurements collected on detector 1 and 2, respectively.  The black and red curves are the predicted transmission corresponding to the light on detector 1 and 2, respectively.}
\end{figure}

\begin{figure}[htbp]
\centering
\includegraphics[width=\linewidth]{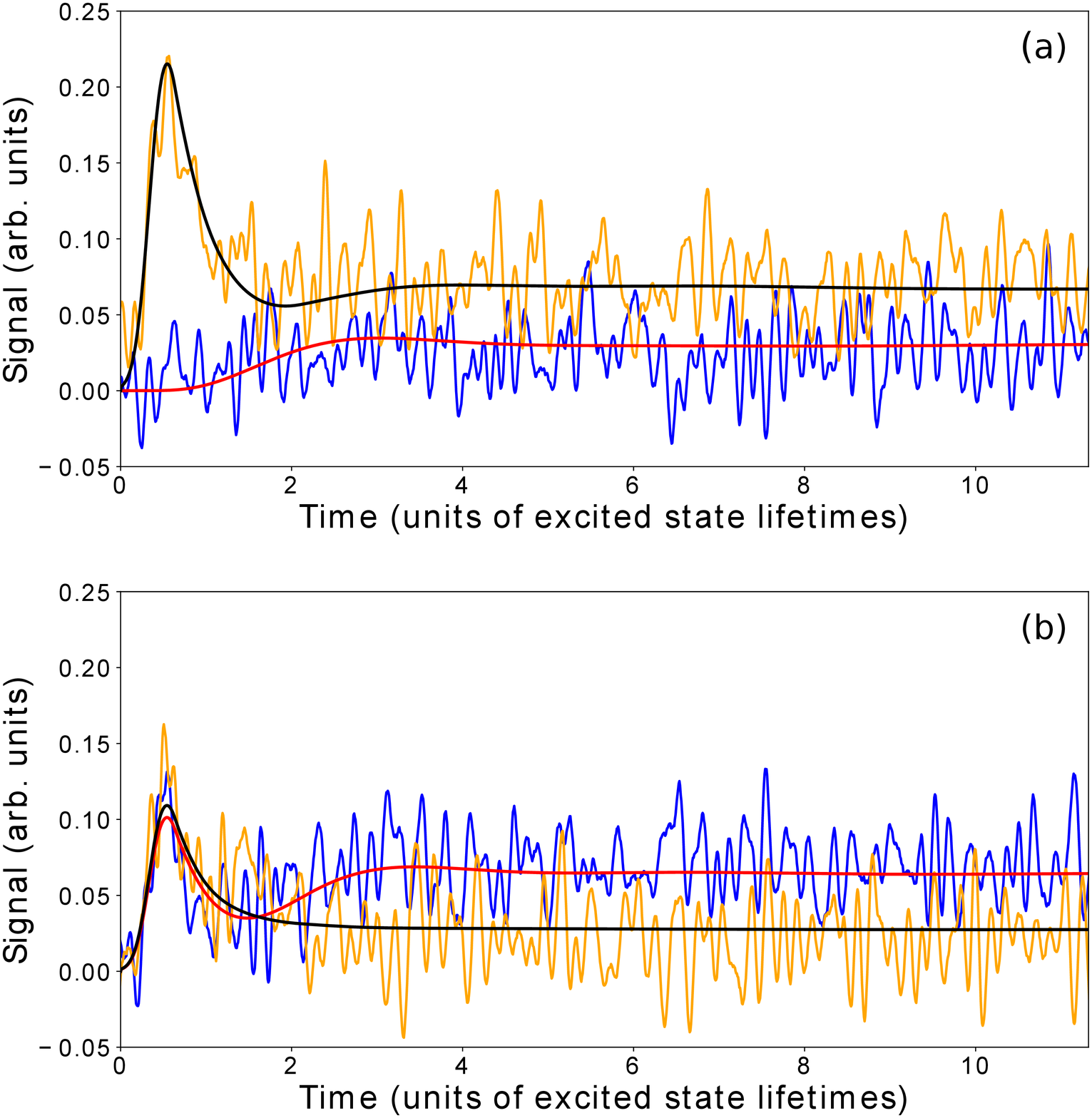}
\caption{\label{fig:Data44} The same type of data as shown in figure \ref{fig:Data13}, except $B = 4.4$ Gauss.}
\end{figure}

\begin{figure}[htbp]
\centering
\includegraphics[width=\linewidth]{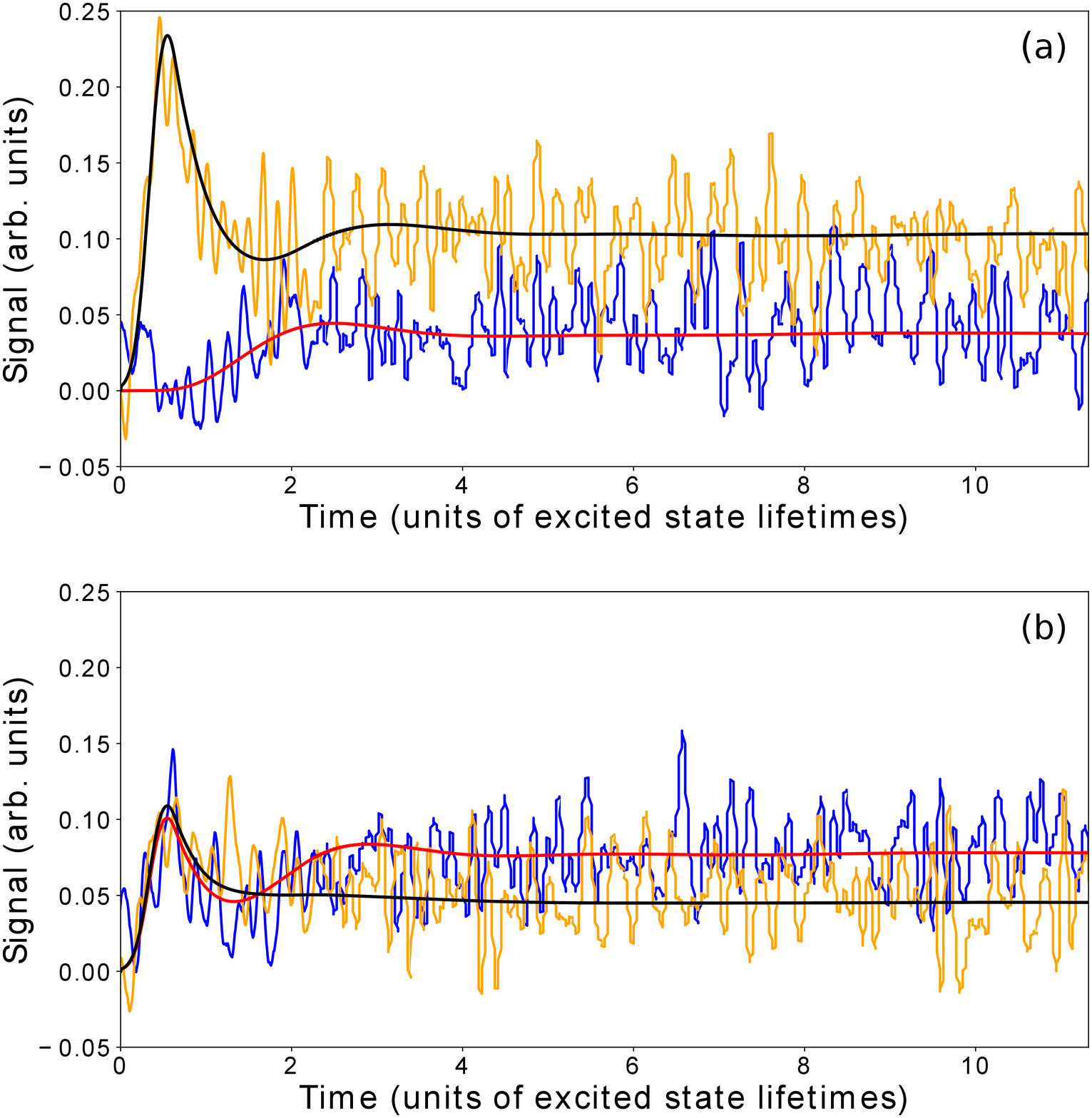}
\caption{\label{fig:Data62} The same type of data as shown in figure \ref{fig:Data13}, except $B = 6.2$ Gauss.}
\end{figure}

	Data collection was performed by interleaving measurements using the 0$^{\circ}$ and 45$^{\circ}$ orientations and two predetermined magnetic field values.  The data collection sequence constrained the parameters in our theory calculations.  Approximately 24 repeated measurements were taken for each specific waveplate and magnetic field combination, and the measurements were combined together to produce curves like those shown in figures \ref{fig:Data13}-\ref{fig:Data62}.  Each of those figures show experimental data collected in the presence of ultracold atoms in the vacuum and corresponding theoretical transmission curves.

	As shown in figures \ref{fig:Data13}-\ref{fig:Data62}, the initial transparency of the gas leads to a peak in the transmission as the incident light turns on, and this intensity peak is visible in data collected using both waveplate orientations.  The initial transparency is a direct result of the atoms requiring a finite amount of time to develop an appreciable dipole amplitude in response to the incident light.  In addition, a time-dependent polarization rotation develops as the atoms impart a differential phase shift between the polarization components of the light.  The signal from this Faraday rotation can be seen in the data collected with detector 2 using the 0$^{\circ}$ waveplate orientation.
	
\section{Time-Dependent Response}
\label{sec:PhaseResponse}

	We have compared our experimental data to theoretical predictions, and good agreement was obtained.  To characterize the time-dependent Faraday effect, we use the phase difference defined by \eqref{eq:Phasediff}.  The phase difference, $\Delta\phi (t)$, can also be extracted through fitting smooth curves directly to the measured data.  Given the agreement between data and theory predictions, however, the basic features of the time-dependent development of the polarization rotation can be extracted from the matched theory curves with little difference from a direct determination from the data. 

\begin{figure}[htbp]
\centering
\includegraphics[width=\linewidth]{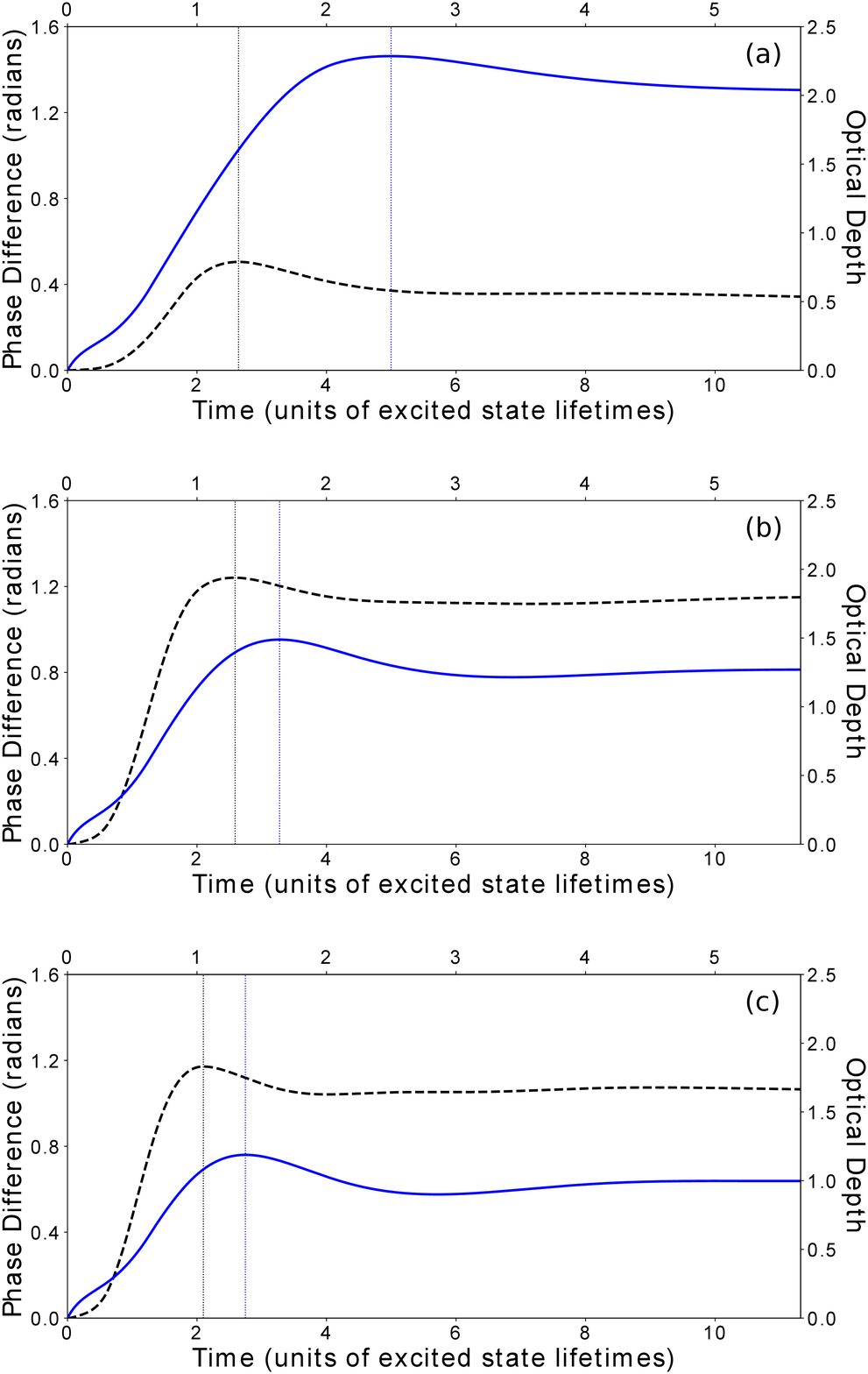}
\caption{\label{fig:PhaseCurves}Time-dependent optical depth (blue solid curve) and time-dependent phase difference (black dashed curve) calculated from the theory curves of figures \ref{fig:Data13}-\ref{fig:Data62}.  The optical depth is plotted with respect to the top horizontal axis and the phase response is plotted with respect to the bottom horizontal axis.  The subfigures (a)-(c) are the results when the magnitude of the magnetic field quantization axis is (a) $B = 1.3$ Gauss, (b) $B = 4.4$ Gauss, and (c) $B = 6.2$ Gauss.  The blue (black) dotted vertical line corresponds to the time when the optical depth (phase response) reaches its maximum.}
\end{figure}

	In addition to determining the phase response, we calculate the time-dependent development of the opacity in the ultracold gas (i.e. the evolution of the optical depth).  This is expressed as 

\begin{equation}
\label{eq:Opacity}
\mathrm{O.D.}(t) = -\ln\left(\frac{I(t)}{I_0(t)}\right), 
\end{equation}	

\noindent
where $I(t)$ is the total transmitted light intensity as a function of time and $I_0(t)$ is the total incident light intensity as a function of time.  The results of the phase response calculations and the evolution of the opacity calculations for data sets collected using three different magnetic fields are shown in figure \ref{fig:PhaseCurves}.
	        	
	The curves shown in figure \ref{fig:PhaseCurves} share expected characteristics with the simple calculations (see section \ref{sec:Simple}) for an ultracold gas of atoms with multiple ground state magnetic sublevels.  These characteristics include a separation between the scaled peak response times of the optical depth and phase difference, along with increasingly faster response times for larger applied magnetic fields.  Figure \ref{fig:PhaseCurves} also illustrates the difference between the time-dependent response of the amplitude component and the time-dependent response of the phase component near $t=0$ for a realistic finite turn-on time.  The amplitude response begins to develop almost immediately after the incident light is turned on, whereas the phase response has a delayed onset under all conditions.  This is expected given the functional time-dependence of the amplitude and the phase components of the light, as discussed in section \ref{sec:Simple}.  

	From the model developed using a multi-level atom gas with no constraints on the incident light intensity, it is straightforward to identify various parameters that can influence the optical depth and phase response times in a detectable way.  These parameters include the incident light frequency detuning and the magnetic field.  For instance, larger magnetic fields have larger detunings and that drives the dipole response to a peak value more rapidly than for smaller magnetic fields.  Additionally, the particular m-state distribution plays a role by weighting the Zeeman induced detuning contributions associated with each transition.  Also, saturation effects influence the response times by making them shorter, although for our conditions the impact of saturation effects on the fitted response times is only a few percent.
	
	As discussed in section \ref{sec:Theory}, the optical thickness plays a role in the gas response time in a way that is different than the factors discussed above.  For an optically thick gas in steady-state, the atoms on the side of the gas opposite to the incident light will have relatively small dipole amplitudes.  This is a result of the light intensity at those atoms' location being less than the incident light due to absorption in the gas \cite{doi:10.1080/09500340.2016.1215564}.  When the light is initially applied to the gas, though, there is much less absorption, and the atoms on the opposite side of the gas are driven by light that has a higher intensity than in steady-state.  This does not persist for long, since the gas absorbs the incident light more and more as a function of time.  However, the larger-than-steady-state intensity drives those atoms toward (or even past) their steady-state dipole response much faster during the period of relative transparency.  This means that the larger the optical thickness of the gas, the shorter the timescale for the gas to absorb the light, since more atoms will be strongly ``overdriven" at early times.  The result is a total transmitted intensity that has a characteristic timescale that is faster than the excited state lifetime of the atoms. 
	
	To illustrate this optical thickness effect with an example, we can compare a couple of data sets with a $B = 1.3$ Gauss magnetic field (where detuning effects were smallest) using gases having an optical depth of $\mathrm{O.D.}=1.3$ and $\mathrm{O.D.}=2.2$.  The decay time of the transmitted intensity signal can be evaluated using the total transmission observed (sum of detector 1 and detector 2 signal) with the waveplate in the 0$^{\circ}$ orientation.  From this total signal, the response time of the gas can be characterized by the time difference between when the transmission peak occurs and the time when the transmission signal has fallen to $e^{-1}$ of the peak value (adjusted by the steady-state transmission signal).  The timescale determined from the $\mathrm{O.D.}=1.3$ data is 0.56 excited state lifetimes and the timescale determined from the $\mathrm{O.D.}=2.2$ data is 0.44 excited state lifetimes illustrating a faster response with increasing optical depth.       

	A similar effect occurs for the development of polarization rotation with time as a function of the optical thickness of the gas, but the effect is not as pronounced as in the absorption case.  As discussed above, the phase response of the atoms is slower than the absorption response and so the polarization component overdrive is reduced as the intensity falls relatively faster than the phase response.  In addition, at early times (after the incident light turn-on) the atoms are being driven by light with a phase that is different than in steady-state.  This tends to mute the polarization overdrive effect to a greater extent than the absorption effect.  Nevertheless, a reduction in the time to reach steady-state as much as a factor of 2 is not uncommon for our conditions, and so this is still a significant effect.
	 
\section{Conclusions}
\label{sec:Conclusions}
	
	Atoms in an ultracold gas do not respond instantaneously to the sudden application of near-resonant light, and this has important consequences for the underlying dynamics in the system before it reaches steady-state.  We have theoretically described and experimentally measured the time-dependent development of Faraday rotation in an ultracold gas subjected to an applied magnetic field, and good agreement between experimental results and theoretical predictions was obtained.  Polarization rotation is ultimately due to phase shifts induced by the real part of indices of refraction in the gas, and a naive expectation would be that the development of those phase shifts requires about a factor of 2 longer in time than absorption effects.  The actual situation is more complex and involves numerous factors that influence the phase response timescale in ways that make general characterizations difficult.  For realistic systems, the phase response is slower than the absorption response, though.  
	
	Among the various parameters of the system that influence the response times is the optical thickness of the gas.  During the initial application of a light pulse, the atoms on the opposite side of the gas from the incident light are driven much more rapidly towards their steady-state response than they would be in an optically thinner gas.  Capturing this optical thickness effect is important for accurately describing the development of absorption and phase responses in an ultracold gas, and would likely need to be considered in any applications or experiments using very short near-resonant light pulses in similar systems.   

\bibliography{References}

\end{document}